\documentclass[twocolumn,prl,showpacs,floatfix]{revtex4}

\usepackage{graphicx}
\usepackage{dcolumn}
\usepackage{bm}
\usepackage{amsmath}

\begin{document}

\title{
Electron localization and entanglement in a two-electron quantum dot
}

\author{Constantine Yannouleas}
\author{Uzi Landman}

\affiliation{School of Physics, Georgia Institute of Technology,
             Atlanta, Georgia 30332-0430}

\date{25 January 2005}

\begin{abstract}
Calculations for two electrons in an elliptic quantum dot,
using symmetry breaking at the unrestricted Hartree-Fock level and subsequent 
restoration of the broken parity via projection techniques, show that the 
electrons can localize and form a molecular dimer, described by a 
Heitler-London-type wave function. The calculated singlet-triplet splitting 
($J$) as a function of the magnetic field ($B$) agrees with cotunneling 
measurements. Knowledge of the dot shape and of $J(B)$ allows determination of 
the degree of entanglement in the ground state of the dot, which is of interest 
for the implementation of quantum logic gates. The theoretical value agrees with 
the experimental estimates.  
\end{abstract}


\maketitle

Electron localization leading to formation of molecular-like structures 
[so-called Wigner molecules (WMs)] within a {\it single circular\/} 
two-dimensional (2D) quantum dot (QD) at zero magnetic field ($B$) has been
theoretically predicted to occur \cite{yl1,yl2p}, as the strength of the 
interaction relative to the zero-point energy increases.
Of particular interest is a two-electron $(2e)$ WM, in light of the
proposal \cite{burk} for the implementation of a solid-state quantum logic
gate that employs two coupled one-electron QDs (double dot). 

In currently fabricated circular QDs the strength of the effective
Coulomb repulsion is significantly reduced (compared to the value used in
Refs.\ \cite{yl1,yl2p}, appropriate for bulk GaAs) due to screening from the 
electrostatic gates and the influence of the finite height of the dot 
\cite{kou2,kyri}. However, changing the shape of the QD (from a circular shape 
to an elliptical one, and ultimately to a quasi-linear one) will reduce the 
zero-point energy, thus enhancing the relative importance of the Coulomb 
repulsion; this assists in bringing the QD into the strongly correlated regime 
resulting in electron localization. Here we show that 
this theoretical prediction \cite{yl3} has indeed been recently observed 
experimentally for two electrons in an elliptic lateral QD \cite{marc}. We 
present microscopic calculations for two electrons in an elliptic QD 
specified by the parameters of the experimental device \cite{marc}.
These calculations show formation of an electron molecular dimer and yield
good agreement with the measured $J(B)$ curve (the singlet-triplet splitting) 
when the value of the Coulomb repulsion is weakened (by 40\%).

Of special interest for quantum computing is the degree of entanglement 
exhibited by the two-electron molecule in its singlet state \cite{burk}. 
A measure of entanglement, introduced in Ref.\ \cite{schl}, 
is known as the concurrence. To date, however, applications \cite{schl,loss} of
this measure have been restricted to the specific singlet state associated
with the {\it bonding\/} and {\it antibonding\/} orbitals of a 
$2e$ {\it double dot\/} with weak interdot-tunneling coupling.
We show that our {\it wave-function-based\/} method enables 
calculation of the concurrence of the $2e$ singlet state in more general cases, 
and in particular for the {\it single\/} elliptic QD of Ref.\ \cite{marc}. This 
is based on our finding that the concurrence of the singlet state is directly 
related to the degree of parity breaking present in the unrestricted 
Hartree-Fock (UHF) orbitals (see below) resulting from our calculations.

We show that knowledge of the dot shape and the $J(B)$ curve allows 
theoretical determination of the degree of entanglement. This supports the 
experimental assertion \cite{marc} that cotunneling spectroscopy can probe 
properties of the electronic wave function of the QD, and not merely 
its low-energy spectrum. 

The hamiltonian for two 2D interacting electrons is
\begin{equation}
{\cal H} = H({\bf r}_1)+H({\bf r}_2)+e^2/(\kappa r_{12}),
\label{ham}
\end{equation}
where the last term is the Coulomb repulsion, $\kappa$ is the
dielectric constant, and 
$r_{12} = |{\bf r}_1 - {\bf r}_2|$. $H({\bf r})$ is the
single-particle hamiltonian for an electron in an external perpendicular
magnetic field ${\bf B}$ and an appropriate confinement potential.
For an elliptic QD, the single-particle hamiltonian is written as
\begin{equation}
H({\bf r}) = T + \frac{1}{2} m^* (\omega^2_{x} x^2 + \omega^2_{y} y^2)
    + \frac{g^* \mu_B}{\hbar} {\bf B \cdot s},
\label{hsp}
\end{equation}
where $T=({\bf p}-e{\bf A}/c)^2/2m^*$, with ${\bf A}=0.5(-By,Bx,0)$ being the
vector potential in the symmetric gauge. $m^*$ is the effective mass and
${\bf p}$ is the linear momentum of the electron. The 
last term in Eq. (\ref{hsp}) is the Zeeman interaction with $g^*$ being the
effective $g$ factor, $\mu_B$ the Bohr magneton, and ${\bf s}$ the spin
of an individual electron. 

Our method for solving the two-body problem defined by the hamiltoninian 
(\ref{ham}) consists of two steps. 
In the first step, we solve selfconsistently the ensuing 
unrestricted Hartree-Fock (UHF) equations allowing for breaking of the total
spin [and {\it parity\/} along the long axis ($x$-axis) of the elliptic QD]. 
For the $S_z=0$ solution, this step produces two single-electron 
orbitals $u_{L,R}({\bf r})$ that are localized left $(L)$ and right $(R)$ of the
center of the QD. At this step, the many-body wave function is a single Slater 
determinant $\Psi_{\text{UHF}} (1\uparrow,2\downarrow) \equiv 
| u_L(1\uparrow)u_R(2\downarrow) \rangle$ made out of the two occupied UHF 
spin-orbitals $u_L(1\uparrow) \equiv u_L({\bf r}_1)\alpha(1)$ and 
$u_R(2\downarrow) \equiv u_R({\bf r}_2) \beta(2)$, where 
$\alpha (\beta)$ denotes the up (down) [$\uparrow (\downarrow)$] spin. 
This UHF determinant is an eigenfunction of the projection $S_z$ of the total 
spin ${\bf S} = {\bf s}_1 + {\bf s}_2$, but not of ${\bf S}^2$ (or the parity
space-reflection operator). 

In the second step, we restore the broken parity and total-spin symmetries by 
applying to the UHF determinant the projection operator \cite{yl3} 
$P^{s,t}=1 \mp \varpi_{12}$, where the operator $\varpi_{12}$ interchanges 
the spins of the two electrons; the minus sign corresponds to the singlet. 
The final result is a generalized Heitler-London (GHL) two-electron wave function
$\Psi^{s,t}_{\text{GHL}} ({\bf r}_1, {\bf r}_2)$ for the ground-state singlet 
(index $s$) and first-excited triplet (index $t$), which uses
the UHF localized orbitals,
\begin{equation}
\Psi^{s,t}_{\text{GHL}} ({\bf r}_1, {\bf r}_2) \propto
\biglb( u_L({\bf r}_1) u_R({\bf r}_2) \pm u_L({\bf r}_2) u_R({\bf r}_1) \bigrb)
\chi^{s,t},
\label{wfghl}
\end{equation}
where $\chi^{s,t} = (\alpha(1) \beta(2) \mp \alpha(2) \beta(1))$ is the spin 
function for the 2$e$ singlet and triplet states.

The use of {\it optimized\/} UHF orbitals in the GHL is suitable for treating 
{\it single elongated\/} QDs. The GHL is equally applicable to double QDs with 
arbitrary interdot-tunneling coupling \cite{yl3}. In contrast,
the Heitler-London (HL) treatment \cite{hl} (known also as Valence bond), 
where non-optimized ``atomic'' orbitals of two isolated QDs are used, is 
appropriate only for the case of a double dot with small interdot-tunneling 
coupling \cite{burk,note1}.

\begin{figure}[t]
\centering\includegraphics[width=7.0cm]{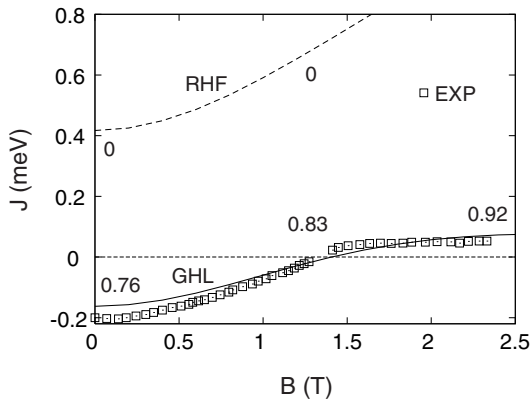}
\caption{
The singlet-triplet splitting $J=E^s-E^t$ as a function of the magnetic field
for an elliptic QD with $\hbar \omega_x=1.2$ meV and $\hbar \omega_y=3.3$ meV
(these values correspond to the device of Ref.\ \cite{marc}). Lower (solid)
curve: The GHL (broken-symmetry UHF + restoration of symmetries) result. 
Upper (dashed) curve: The restricted HF (RHF, no symmetry breaking) result.
The experimental measurements are denoted by the open squares \cite{marc}.
The material parameters used are: $m^*$(GaAs)$=0.067 m_e$, $\kappa=22.0$,
and $g^*=0$ (see text). The calculated values for the concurrence of the
singlet GHL state (${\cal C}^s$, see text) at $B=0$, 1.3 T, and 2.5 T
are also displayed. In the singlet RHF case (upper curve), the concurrence
is identically zero for all values of $B$. Note that our sign convention for $J$
is opposite to that in Ref.\ \cite{marc}.
}
\end{figure}

Both the GHL singlet, $\Psi^s_{\text{GHL}}$, and the GHL triplet,
$\Psi^t_{\text{GHL}}$, cannot be reduced to a single Slater determinant. 
They are always the sum of two Slater determinants,
i.e., $\Psi^{s,t}_{\text{GHL}} \propto
|u_L(1\uparrow)u_R(2\downarrow) \rangle \mp 
|u_L(1\downarrow)u_R(2\uparrow) \rangle$,
and thus they represent states with intrinsic entanglement \cite{burk,yl3},
a property that underlies the operation of the quantum logic gate. 

The 2D UHF equations that we are using are described in detail in Ref.\ [1(c)].
For the formulas which enable calculation of the energies $E^{s,t}_{\text{GHL}}$
for the special case of the two-electron wave function (\ref{wfghl}), see 
Refs.\ \cite{yl3,yl4}. 
The key point is that we exploit an additional variational freedom by allowing
for different orbitals [here $u_L({\bf r})$ and $u_R({\bf r})$] for the two spin 
directions (up and down). Under appropriate conditions, the UHF equations do 
indeed have solutions associated with two spatially separated orbitals (symmetry
breaking is present). We note, however, that for sufficiently weak Coulomb 
repulsion (very large dielectric constant $\kappa$, see below), the two 
orbitals $u_L({\bf r})$ and $u_R({\bf r})$ do collapse onto the same single 
orbital $u({\bf r})$, and there is no symmetry breaking.  

Another pertinent point here is that the orbitals $u_{L,R}({\bf r})$ are 
expanded in a real Cartesian harmonic-oscillator basis, namely,
\begin{equation}
u_{L,R}({\bf r}) = \sum_{j=1}^K C_j^{L,R} \varphi_j ({\bf r}),
\label{uexp}
\end{equation}
where the index $j \equiv (m,n)$ and $\varphi_j ({\bf r}) = X_m(x) Y_n(y)$,
with $X_m(Y_n)$ being the eigenfunctions of the one-dimensional oscillator in the
$x$($y$) direction with frequency $\omega_x$($\omega_y$). The parity operator
${\cal P}$ yields ${\cal P} X_m(x) = (-1)^m X_m(x)$, and similarly for $Y_n(y)$.
The expansion coefficients $C_j^{L,R}$ are real for $B=0$ and complex for finite
$B$.

We turn now to the interpretation of the measurements of Ref.\ \cite{marc}. To 
model the experimental elliptic QD device, we take, following Ref.\ \cite{marc},
$\hbar \omega_x=1.2$ meV and $\hbar \omega_y=3.3$ meV. 
The effective mass of the electron is taken as 
$m^*=0.067 m_e$ (GaAs). Since the experiment did not resolve the lifting of
the triplet degeneracy caused by the Zeeman term, we take $g^*=0$.
Using our two step method described above, we calculate the 
singlet-triplet splitting 
$J_{\text{GHL}}(B)=E^s_{\text{GHL}}(B)-E^t_{\text{GHL}}(B)$ 
as a function of the magnetic field in the range $0 \leq B \leq 2.5$ T.
Similar to an earlier study  \cite{kyri}, we find that a weakening of the
Coulomb repulsion from its value in GaAs $(\kappa = 12.9)$ is required in
order to reproduce the experimental $J(B)$ curve. The effect of screening
by the gates can be modeled, to first approximation, by increasing 
$\kappa$ \cite{kyri,note2}. Indeed, with $\kappa=22.0$, very good agreement with
the experimental data (see Fig.\ 1) is obtained. In particular, we 
note the singlet-triplet (ST) crossing about 1.3 T, and the backbending of the 
$J(B)$ curve beyond this crossing. 

In Fig.\ 1, we also plot for $\kappa=22.0$ the $J_{\text{RHF}}(B)$ curve [upper 
(dashed) line] obtained from the {\it restricted\/} HF (RHF) calculation, namely
for a self-consistent Hartree-Fock variation with the restriction that the 
{\it parity be preserved\/}, so that there is no symmetry breaking and
both the spin-up and spin-down electrons occupy the same spatial
orbital [i.e., in this approximation $u_L({\bf r})=u_R({\bf r})=u({\bf r})$].
The $J_{\text{RHF}}(B)$ curve fails substantially to reproduce the experimental 
results. The RHF curve disagrees with the experimental trends in 
two important ways: (I) $J_{\text{RHF}}(0)$ is larger than zero, which
not only contradicts the experiment, but also a fundamental theoretical result 
[1(b), 14] that states that for two electrons at zero magnetic field 
the singlet is always the ground state, and (II) The $J_{\text{RHF}}(B)$ curve 
diverges as $B$ increases, unlike the $J_{\text{GHL}}(B)$ curve (as well as the 
experimental observation) which bends back after the ST crossing and approaches 
asymptotically the $J=0$ line.

\begin{figure}[t]
\centering\includegraphics[width=6.5cm]{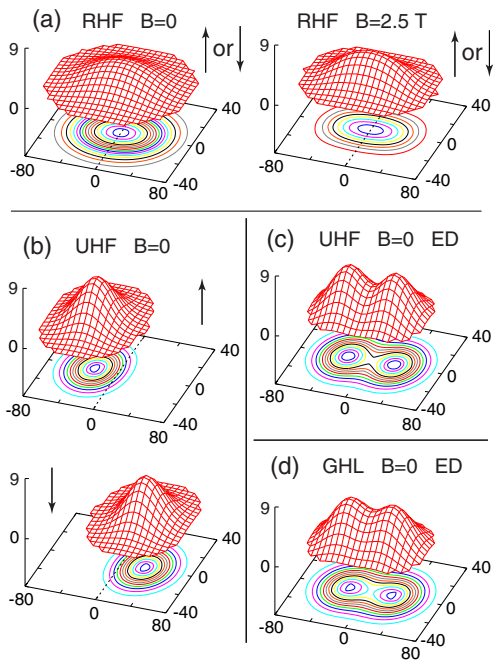}
\caption{
Spin-up and spin-down orbitals (modulus square) and total electron denities 
(EDs) associated with the singlet state of two electrons in the elliptic QD. 
The arrows indicate a spin-up or a spin-down electron. 
(a) Doubly occupied orbitals of the parity-preserving RHF at $B=0$ (left) and 
$B=2.5$ T (right).
(b) Orbitals of the broken-symmetry UHF at $B=0$.
(c-d) EDs at $B=0$ for the UHF (c) and the GHL (d) wave functions.
The GHL ED is larger in the region between the two humps compared to 
the UHF one; this is indicative of covalent bonding.
The rest of the parameters are as in Fig.\ 1.
Lengths in nm and orbital densities in $10^{-4}$ nm$^{-2}$.
}
\end{figure}

The sharp contrast in the behavior of the RHF and GHL results further highlights
the significance of the latter. Indeed the behavior of the RHF solution reflects 
the independent-particle-model nature of the RHF orbitals [see Fig.\ 2(a)], 
which leads to a preponderance of the exchange contribution, and thus to 
spontaneous (but erroneous) magnetization of the electron system. On the other 
hand, the GHL results in Fig.\ 1 reflect the fact that symmetry breaking and 
electron localization [see orbitals in Fig.\ 2(b)] reduce the Coulomb repulsion 
to a degree that compensates for the loss of exchange binding (since the
localization reduces the orbital overlap). This is portrayed in formation of a 
Wigner molecule with the aforementioned agreement with the experimental $J(B)$
curve. The total electron density (ED) of the WM resulting from breaking of
symmetry [UHF, Fig.\ 2(c)] and after restoration of parity (and total-spin)
symmetry [GHL, Fig.\ 2(d)] illustrates the electron molecular dimer.
We note that the UHF solutions
at $B=0$ exhibit breaking of the parity for values of $\kappa$ as large as 
$\kappa=40.0$, which indicates that the value of $\kappa=22.0$ in our 
calculations places the elliptic device of Ref.\ \cite{marc} well within the 
regime of strong electron correlations.

The asymptotic energetic convergence (beyond the ST point) of the singlet and 
triplet states, i.e., [$J(B) \rightarrow 0$ as $B \rightarrow \infty$] is a 
reflection of the dissociation of the 2$e$ molecule, since the ground-state 
energy of two fully spatially separated electrons (zero overlap) does not 
depend on the total spin. Indeed, orbitals and EDs for finite $B$ values are
similar to those in Figs.\ 2(b-d), but with enhanced localization reflected in
diminished overlaps with increasing $B$.

Both the GHL singlet and triplet wave functions
[Eq.\ (\ref{wfghl})] exhibit entanglement by being the sum of two Slater 
determinants \cite{yl3,schl}. However, unlike the triplet GHL state which is 
maximally entangled (see below), the singlet GHL wave function may exhibit
a reduced degree of entanglement, as was found from the Hund-Mulliken
treatment of the double dot \cite{loss}. For purposes of quantum computing, it 
is necessary to be able to extract from this GHL singlet a quantitative measure 
of the degree of entanglement, e.g., the corresponding value of the concurrence 
${\cal C}$ \cite{schl,loss}. The concurrence, however, was introduced for the 
Hilbert space associated with the {\it bonding\/} and {\it antibonding\/} 
orbitals of a double dot molecule \cite{schl,loss}. In our case of a single 
elliptic QD, the concept of a bonding and antibonding orbital does not 
straightforwardly apply. Rather, we need to utilize the
universal symmetry properties of the bonding and antibonding orbitals with 
respect to the parity operator, namely, the bonding orbital is symmetric $(+)$, 
and the antibonding orbital is antisymmetric $(-)$ with respect to reflection 
about the origin of the $x$-axis. We first notice that, due to the 
symmetry breaking, the UHF orbitals $u_L({\bf r})$ and $u_R({\bf r})$ 
are not eigenstates of the parity operator, and we proceed to separate the 
symmetric $\Phi^+({\bf r})$ and antisymmetric $\Phi^-({\bf r})$ components in 
their expansion given by Eq.\ (\ref{uexp}). That is, we write
$u^{L,R}({\bf r}) \propto \Phi^+({\bf r}) \pm  \xi \Phi^-({\bf r})$;
note that $\Phi^+({\bf r})$ and $\Phi^-({\bf r})$ are eigenfunctions of the
parity operator. Subsequently, with the use of Eq.\ (\ref{wfghl}), the GHL 
singlet can be rearranged as follows:
\begin{equation}
\Psi^{s}_{\text{GHL}} \propto
| \Phi^+(1\uparrow) \Phi^+(2\downarrow) \rangle - 
\eta |\Phi^-(1\uparrow)\Phi^-(2\downarrow) \rangle,
\label{rear}
\end{equation}
where the coefficient in front of the second determinant, $\eta=\xi^2$, is the 
so-called interaction parameter \cite{loss}. 
Knowing $\eta$ allows a direct evaluation of the concurrence of the singlet
state, since ${\cal C}^s = 2\eta/(1+\eta^2)$ \cite{loss}. Note that 
$\Phi^+({\bf r})$ and $\Phi^-({\bf r})$ are properly normalized and that they 
are by construction orthogonal. 

For the GHL triplet, one obtains an expression independent of the 
interaction parameter $\eta$, i.e.,
\begin{equation}
\Psi^{t}_{\text{GHL}} \propto
| \Phi^+(1\uparrow) \Phi^-(2\downarrow) \rangle +
|\Phi^+(1\downarrow)\Phi^-(2\uparrow) \rangle,
\label{reart}
\end{equation}
which is a maximally (${\cal C}^t=1$) entangled state. Note that underlying the 
analysis of Ref.\ \cite{marc} is a {\it conjecture\/} that wave functions of the
form given in Eqs.\ (\ref{rear}) and (\ref{reart}) describe the two electrons
in the elliptic QD.

To make things more concrete, we display in TABLE I for $B=0$ (and 
$\kappa=22.0$) the coefficients $C_j^{L,R}$ [see Eq.\ (\ref{uexp})] that specify
the broken-symmetry UHF orbitals in the real Cartesian harmonic-oscillator 
basis. Indeed, we find that there are contributions from both symmetric
and antisymmetric basis functions $X_m(x)Y_n(y)$ along the $x$-axis, since both
even and odd $m$ indices are present. The $n$ indices are all
even, since there is no symmetry breaking along the $y$-axis.
Naturally, all the $m$ indices in the expansion of the RHF common 
$u({\bf r})$ orbital are even, since in this case the parity is preserved.
One can immediately check that 
$ \xi = \left( \sum_{j\;(m\; {\text{odd}})}^K |C_j^L|^2  \left/ 
\sum_{j\;(m\; {\text{even}})}^K |C_j^L|^2 \right. \right)^{1/2}$.

\begin{table}[t]
\caption{%
Left: Expansion coefficients $C_j^{L,R}$ [See Eq.\ (\ref{uexp})] for the
broken-symmetry UHF orbitals $u_L({\bf r})$ and $u_R({\bf r})$ at $B=0$ and
$\kappa=22.0$. Right: Expansion coefficients for the RHF common orbital 
$u({\bf r})$. The real-Cartesian-harmonic-oscillator basis functions are given by
$\varphi_j ({\bf r}) = X_m(x) Y_n(y)$. The running index $j$ stands for
the pair $(m,n)$, where $m$ and $n$ denote the number of nodes along the
$x$ and $y$ directions, respectively. A total of $K=79$ basis states 
$\varphi_j ({\bf r})$ were used. However, only coefficients with absolute value
larger than 0.01 are listed here. For numbers with a $\pm$ in the front, the
$+$ corresponds to the left $(L)$ orbital and the $-$ to the right $(R)$ one.
}
\begin{ruledtabular}
\begin{tabular}{cdcd} 
\multicolumn{2}{c}{UHF}& \multicolumn{2}{c}{RHF} \\ \hline
\multicolumn{1}{c}{$j~(m,n)$}   & 
\multicolumn{1}{c}{$~~~~~C_j^{L,R}$} & 
\multicolumn{1}{c}{$j~(m,n)$}   & 
\multicolumn{1}{c}{$~~~~~~C_j$}       \\ \hline 
~1~(0,0)  &        0.7990  & ~1~(0,0)  & 0.9710 \\ 
~2~(1,0)  &     \pm0.5600  & ~3~(2,0)  & 0.2310 \\
~3~(2,0)  &        0.2150  & ~7~(4,0)  & 0.0172 \\
~5~(3,0)  &     \pm0.0104  & 10~(0,2)  & 0.0613 \\
~7~(4,0)  &       -0.0180  &  ~~~      & ~~~       \\
10~(0,2)  &        0.0253  &  ~~~      & ~~~       \\ 
\end{tabular}
\end{ruledtabular}
\end{table}

For the RHF singlet $\xi=0$, since there no coefficients with odd $m$
indices (TABLE I). For the GHL singlet, we calculate the interaction parameter 
$\eta=\xi^2$ and the concurrence ${\cal C}^s$ for the device of 
Ref.\ \cite{marc} in the magnetic-field range $0 \leq B \leq 2.5$ T. We find
that starting with $\eta=0.46$ $({\cal C}^s=0.76)$ at $B=0$, the interaction
parameter (singlet-state concurrence) increases monotonically to $\eta=0.65$ 
$({\cal C}^s=0.92)$ at $B=2.5$ T. At the intermediate value corresponding to the
ST transition ($B=1.3$ T), we find $\eta=0.54$ $({\cal C}^s=0.83)$. Our $B=0$ 
theoretical result for $\eta$ and ${\cal C}^s$ are in remarkable agreement with 
the experimental estimates \cite{marc} of $\eta=0.5 \pm 0.1$ and 
${\cal C}^s=0.8$, which were based solely on conductance measurements below the 
ST transition (i.e., near $B=0$). 
Our theoretical values for the singlet-state concurrence are marked
on the GHL curve in Fig.\ 1. We also display in this figure the value of
${\cal C}^s$ for the RHF single-determinant solution (upper curve). 
The RHF value for this quantity (and also for $\eta$) vanishes identically for 
all $B$ values. Thus the shape (and  
values) of the experimental $J(B)$ curve portray directly the degree of 
electron localization and the entanglement associated with the singlet state 
in Eq.\ (\ref{wfghl}). 
Note that our calculated results are based 
on the energetics of the singlet-triplet splitting alone, and thus they provide 
a reliable and independent alternative method for extracting the degree of 
singlet-state entanglement for all values of $B$. 

In conclusion, we have shown formation of an electron molecular dimer in an
elliptic QD (Fig.\ 2) for screened interelectron repulsion characterized by
a singlet-triplet splitting $J(B)$ that agrees with experiment (Fig.\ 1).
Furthermore we showed that, from a knowledge of the dot shape and of $J(B)$, 
theoretical analysis along the lines introduced here allows probing of the 
correlated ground-state wave function and determination of its degree of 
entanglement. Such information is of great value for the implementation of 
solid-state quantum logic gates. It is also of interest to quantum information 
theory in general \cite{vers,pres}. The present theoretical method and analysis 
can be straightforwardly extended to double dots with arbitrary 
interdot-tunneling coupling.

This research is supported by the US D.O.E. (Grant No. FG05-86ER45234), and
NSF (Grant No. DMR0205328). We thank M. Pustilnik for comments on the
manuscript.


\begin{thebibliography}{99}
\bibitem{note1}
An immediate extension of the simple HL method known
as the Hund-Mulliken method is also restricted to the case
of a double QD with weak interdot tunneling.
\end{thebibliography}
\end{document}